\documentclass[aps,prb,showpacs,twocolumn,floats]{revtex4}
\usepackage{subfigure}
\usepackage{epsfig}
\usepackage{dcolumn}
\usepackage{bm}
\usepackage[ansinew]{inputenc}
\usepackage{amsmath}
\usepackage{amsthm}
\usepackage[T1]{fontenc}
\usepackage{amssymb}
\usepackage{amsfonts}
\usepackage[english]{babel}
\usepackage{enumitem}
\newcommand{\ud}{\,\mathrm{d}}
\newcommand{\ds}{\displaystyle}

\begin{document}

\title{Dissipative Macroscopic Quantum Tunneling in Type-I Superconductors}
\date{\today}
\author{R. Zarzuela$^{1}$, E. M. Chudnovsky$^{2,1}$, and J. Tejada$^{1}$}
\affiliation{$^{1}$Departament de F\'{i}sica Fonamental, Facultat
de F\'{i}sica, Universitat de Barcelona, Avinguda Diagonal 645,
08028 Barcelona, Spain\\ $^{2}$Physics Department, Lehman College,
The City University of New York, 250 Bedford Park Boulevard West,
Bronx, NY 10468-1589, U.S.A.}
\date{\today}

\begin{abstract}
We study macroscopic quantum tunneling of interfaces separating
normal and superconducting regions in type-I superconductors.
Mathematical model is developed, that describes dissipative
quantum escape of a two-dimensional manifold from a planar
potential well. It corresponds to, e.g., a current-driven quantum
depinning of the interface from a grain boundary or from
artificially manufactured pinning layer. Effective action is
derived and instantons of the equations of motion are
investigated. Crossover between thermal activation and quantum
tunneling is studied and the crossover temperature is computed.
Our results, together with recent observation of non-thermal
low-temperature magnetic relaxation in lead, suggest possibility
of a controlled measurement of quantum depinning of the interface
in a type-I superconductor.
\end{abstract}

 \maketitle

\maketitle
\section{Introduction}

Macroscopic quantum tunneling refers to the situation when an
object consisting of many degrees of freedom, coupled to a
dissipative environment, escapes from a metastable well via
underbarrier quantum tunneling \cite{Caldeira-Leggett}. In
condensed matter this phenomenon was first observed through
measurements of tunneling of the macroscopic magnetic flux created
by a superconducting current in a circuit interrupted by a
Josephson junction \cite{Clarke-88}. Another example is tunneling
of magnetization in solids \cite{CT-book}. In cases of the
magnetic flux or the magnetic moment of a nanoparticle, the
tunneling object is described by one or two macroscopic
coordinates that depend on time, like in a problem of a tunneling
particle in quantum mechanics. The environment enters the problem
through interaction of these macroscopic coordinates with
microscopic excitations of the medium. Equally interesting, but
significantly more involved, is the problem of tunneling of a
macroscopic field between two distinct configurations. Examples
are tunneling of vortex lines in type-II superconductors
\cite{Larkin-review1994,Yeshurun-review1996,Nattermann-review2000}
and tunneling of domain walls in magnets
\cite{Stamp,Giordano,Rosenbaum}. The essential difference between
the last two examples is that tunneling of vortex lines is
determined by their predominantly dissipative dynamics
\cite{Blatter-1991,Ivlev-1991,Tejada-1993,Ao-1994,Stephen-1994},
while tunneling of the spin-field is affected by dissipation to a
much lesser degree. Theory that describes quantum tunneling of
extended condensed-matter objects involves space-time instantons
that are similar to the instantons studied in relativistic field
models. Examples that are available for experimental studies are
limited. Consequently, any new example of tunneling of an extended
object must be of significant interest.

Recent measurements of low-temperature magnetic relaxation of lead
\cite{Pb-tunneling} have elucidated the possibility of macroscopic
quantum tunneling in type-I superconductors. Such superconductors
(with lead being a prototypical system), unlike type-II
superconductors, do not develop vortex lines when placed in the
magnetic field. Instead, they exhibit intermediate state in which
the sample splits into normal and superconducting regions
separated by planar interfaces of positive energy
\cite{Landau,Sharvin,Huebener}. Equilibrium states and dynamics of
interfaces have been well studied by now
\cite{Kuznetsov-1998,Cebers-2005,Menghini-2005,Prozorov-PRL2007,Prozorov-Nature2008,Velez-PRB2009}.
In all these studies the interface was treated as a classical
object. Recently, however, it was noticed \cite{Pb-tunneling} that
slow temporal evolution of magnetization in a superconducting Pb
sample was independent of temperature below a few kelvin. This
observation pointed towards possibility of quantum tunneling of
interfaces in the potential landscape determined by pinning. In
general the pinning potential would be due to random distribution
of pinning centers or due to properties of the sample surface. In
a polycrystalline sample it may also be due to extended pinning of
interfaces by grain boundaries.

\begin{figure}
\includegraphics[width=80mm]{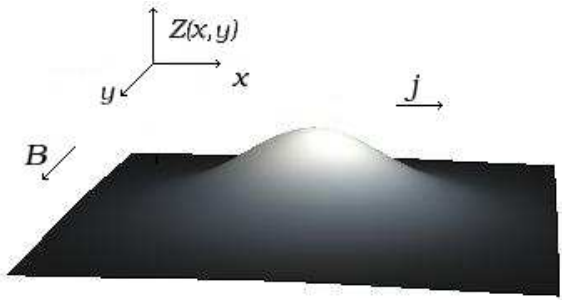}
\vspace{1mm} \caption{Interface between normal and superconducting
regions in a type-I superconductor, pinned by a planar defect in
the XY plane. Transport current parallel to the interface controls
the energy barrier.  Depinning of the interface occurs through
quantum nucleation of a critical bump described by the instanton
of the equations of motion in 2+1 dimensions.} \label{interface}
\end{figure}

Modern atomic deposition techniques permit preparation of a
pinning layer with controlled properties. This inspired us to
study a well defined problem in which the interface separating
normal and superconducting regions is pinned by a planar defect.
The corresponding pinning barrier can be controlled by a
superconducting current that exerts a force on the interface. At
low temperature the depinning of the interface would occur through
quantum nucleation of a critical bump shown in Fig.\
\ref{interface}.
Somewhat similar problems in 1+1 dimensions have been studied for
a flux line pinned by the interlayer atomic potential in a layered
superconductor \cite{Ivlev-1991} and for a flux line pinned by a
columnar defect \cite{CD-1995}. However, the two-dimensional
nature of the interface, as compared to a one-dimensional flux
line, makes the interface problem more challenging. Note that
tunneling of two-dimensional objects has been studied
theoretically in application to non-thermal dynamics of planar
domain walls \cite{Stamp} and quantum nucleation of magnetic
bubbles \cite{bubbles}. These studies employed non-dissipative
dynamics of the magnetization field because corrections coming
from dissipation are not dominant for spin systems. On the
contrary, the Euclidean dynamics of the interface in a type-I
superconductor is entirely dissipative, described by
integro-differential equations in 2+1 dimensions. As far as we
know this problem has not been studied before.

The article is structured as follows. Theoretical model is
formulated in Sec. \ref{model}. Properties of the pinning
potential and the effective action in the vicinity of the critical
depinning current are analyzed in Sec. \ref{Potential}. Instantons
of the dissipative model in 2+1 dimensions are investigated in
Sec. \ref{instanton}. Crossover from quantum tunneling to thermal
activation  is studied in Sec. \ref{crossover}. Sec.
\ref{discussion} contains estimates of the effect and final
conclusions.

\section{The Model}\label{model}
We describe the interface by a smooth function $Z(x,y)$, see Fig.\
\ref{interface}. Dimensionless Euclidean effective action
associated with the interface is
\begin{eqnarray}\label{eq1.1}
&&S_{eff}=\frac{\sigma}{\hbar}\oint\ud \tau\int\ud x\ud y
\left[1+({\bm \nabla}
Z)^2\right]^{\frac{1}{2}} \nonumber \\
&&+\frac{1}{\hbar}\oint\ud \tau\int\ud x\ud y\,
V\left[x,y,Z(x,y,\tau)\right]
\\ && +\frac{\eta}{4\pi\hbar}\oint\ud \tau\int_{\mathbf{R}}\ud
\tau'\int\ud x\ud y\frac{\left[Z(x,y,\tau)-Z(x,y,\tau')\right]^2}
{(\tau-\tau')^2} \nonumber
\end{eqnarray}
where $\tau$ is the imaginary time, $\sigma$ is the surface energy
density of the interface and $\eta$ is a drag coefficient, given
respectively by \cite{LP,Pb-tunneling}
\begin{equation}\label{sigma-eta}
\sigma=\frac{B_c^2\xi}{3\sqrt{2}\pi}\,, \quad \eta =
\frac{B_c^2\sqrt{\lambda_L\xi}}{2\rho_n c^2}\,,
\end{equation}
with $B_c$ being the thermodynamic critical field, $\xi$ being the
superconducting coherence length, $\lambda_L$ being the London
length, and $\rho_n$ being the normal state resistivity. The first
term in Eq.\ (\ref{eq1.1}) is due to the elastic energy of the
interface associated with its total area, the second term is due
to the space-dependent potential energy,
$V\left[x,y,Z(x,y)\right]$, of the interface inside the imperfect
crystal, and the third term is due to dissipation
\cite{Caldeira-Leggett}. Same as for the flux lines, we neglect
the inertial mass of the interface. Its dynamics in a type-I
superconductor is dominated by friction.

We consider pinning of the interface by a planar defect located in
the $XY$ plane and choose the pinning potential in the form
\begin{equation}\label{eq1.2}
V_p = p\sigma\int\ud x\ud y
\;\left(\frac{1}{2}\frac{Z^2}{a^2}-\frac{1}{4}\frac{Z^4}{a^4}\right)
\end{equation}
where $2a$ is roughly the width of the well that traps the
interface and $p \lesssim 1 $ is a dimensionless constant
describing the strength of the pinning. The interface separates
the normal state at $Z < 0$ from a superconducting state at $Z >
0$. Superconducting current parallel to the planar defect (and to
the interface pinned by the defect) exerts a Lorentz force on the
interface similar to the force acting on a vortex line in a
type-II superconductor. We shall assume that the magnetic field is
applied in the $\hat{y}$ direction and that the transport current
of density $j$ flows in the $\hat{x}$ direction. The driving force
experienced by the $\mathrm{d}x\mathrm{d}y$ element of the
interface in the $\hat{z}$ direction is given by
\begin{equation}
\frac{\mathrm{d}^{2}F_{z}}{\mathrm{d}x\mathrm{d}y}=\frac{1}{c}\int\ud
z j B(z)\,,
\end{equation}
Here $B(z)=B_{c}\exp(-z/\delta)$ is the magnetic field inside the
interface with $\delta=\sqrt{\xi\lambda_{L}}$. Integration then
gives ${\mathrm{d}^{2}F_{z}}/({\mathrm{d}x\mathrm{d}y})=
{B_{c}\delta}j/c$. The corresponding contribution to the potential
can be obtained by writing $F_{z}$ as $-\nabla_{Z}V_{L}$, yielding
\begin{equation}
\frac{\mathrm{d}^{2}V_{L}(Z)}{\mathrm{d}x\mathrm{d}y}=-\frac{B_{c}\delta}{c}jZ\,.
\end{equation}
The total potential, $V(Z)=V_{p}(Z)+V_{L}(Z)$ is
\begin{equation}\label{V}
V(Z)=p\sigma \int\ud x\ud y\left(-\bar{j}
\tilde{Z}+\frac{\tilde{Z}^2}{2}-\frac{\tilde{Z}^4}{4}\right)
\end{equation}
where we have introduced dimensionless $\tilde{Z}=Z/a$ and
\begin{equation}
\bar{j}=\frac{a\delta B_c}{p c \sigma}j =
\frac{3\pi\sqrt{2{\kappa}}a}{p c B_c}j
\end{equation}
with $\kappa = \lambda_L/\xi$. Note that for a type-I
superconductor $\kappa < 1/\sqrt{2}$.

\section{Effective Action in the Vicinity of the Critical Current}\label{Potential}

Measurable quantum depinning of the interface can occur only when
the transport current is close to the critical current, $j_c$,
that destroys the energy barrier. It is, therefore, makes sense to
study the problem at $j \rightarrow j_c$. Maxima and minima of the
function
\begin{equation}\label{f}
f(\bar{j},\tilde{Z})=-\bar{j}\tilde{Z}+\frac{\tilde{Z}^2}{2}-\frac{\tilde{Z}^4}{4}
\end{equation}
that enters Eq.\ (\ref{V}) are given by the roots of the equation
$\tilde{Z}^3-\tilde{Z}+\bar{j}=0$. At $j^2 < 4/27$ it has three
real roots corresponding to one minimum and two maxima of the
potential on two sides of the pinning layer, whereas at $j^2 >
4/27$ there is one real root corresponing to the maximum of $f$.
Consequently, the barrier disappears at $j^2 = 4/27$, providing
the value of the critical current
\begin{equation}
\bar{j}_c = \frac{2}{3\sqrt{3}}\,, \quad j_c =
\frac{2pcB_c}{9\pi\sqrt{6\kappa}a}\,.
\end{equation}

At $\bar{j} = \bar{j}_c$ the minimum and the maximum of the
potential combine into the inflection point $\tilde{Z} =
\tilde{Z}_c$ given by the set of equations
\begin{eqnarray}
0&=&-\tilde{Z}_c^3+\tilde{Z}_c-\bar{j}_c \nonumber \\
0&=&-3\tilde{Z}_c^2+1
\end{eqnarray}
that correspond to zero first and second derivatives of $f$. The
value of $\tilde{Z}_c$ deduced from these equations is
${1}/{\sqrt{3}}$. It is convenient to introduce small parameter
\begin{equation}
\epsilon= 1-{j}/{j_c}\,,
\end{equation}
so that $j=j_c(1-\epsilon)$ and
\begin{equation}
\bar{j}=\bar{j}_c(1-\epsilon)=\frac{2}{3\sqrt{3}}(1-\epsilon)\,.
\end{equation}

\begin{figure}
\vspace{-2mm}
\includegraphics[width=97mm]{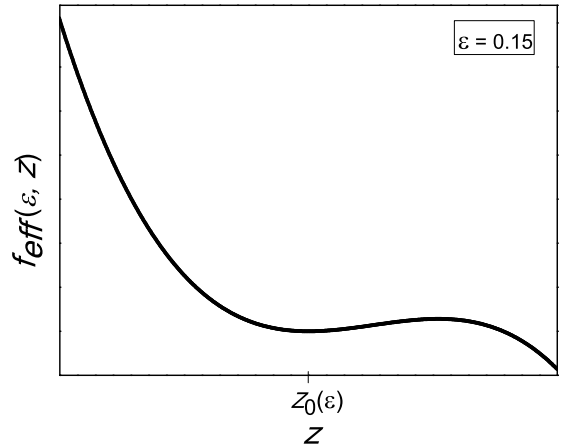}
\caption{Effective potential} \label{potential}
\end{figure}

Let $\tilde{Z}_0(\bar{j})$ be the minimum of $f$ (see Fig.\
\ref{potential}) satisfying
\begin{equation}\label{Z_0}
\tilde{Z}_0^3-\tilde{Z}_0+\bar{j}_c(1-\epsilon)=0\,.
\end{equation}
Consider $\tilde{Z}'=\tilde{Z} - \tilde{Z}_0$. It is easy to find
that the form of the potential in the vicinity of $\tilde{Z}_0$ is
\begin{equation}\label{f-shift}
f = f[\tilde{Z}_0(\bar{j})] +
\frac{1}{2}(1-3\tilde{Z}_0^2)\tilde{Z}'^2-\tilde{Z}_0\tilde{Z}'^3-\frac{\tilde{Z}'^4}{4}\,.
\end{equation}
At small $\epsilon$ one has $\tilde{Z}_0\rightarrow\tilde{Z}_c =
1/\sqrt{3}$, so that $1-3\tilde{Z}_0^2$ in front of $\tilde{Z}'^2$
in Eq.\ (\ref{f-shift}) is small. The first term in Eq.\
(\ref{f-shift}) can be omitted as unessential shift of energy,
while the last term proportional to $\tilde{Z}'^4$ can be
neglected due to its smallness compared to other
$\tilde{Z}'$-dependent terms. Consequently, one obtains the
``effective potential''
\begin{equation}
f_{eff}(\bar{j},\tilde{Z})=\frac{1}{2}(1-3\tilde{Z}_0^2)\tilde{Z}^2-\tilde{Z}_0\tilde{Z}^3\,.
\end{equation}
We need to know the dependence of $\tilde{Z}_0$ on $\epsilon$.
Writing $\tilde{Z}_0(\epsilon)=\tilde{Z}_c[1-\beta(\epsilon)]$,
with the help of Eq.\ (\ref{Z_0}), we obtain $\ds \beta(\epsilon)
=\sqrt{2\epsilon/3}$ to the lowest order on $\epsilon$. Then
$1-3\tilde{Z}_0^2\approx2\sqrt{2\epsilon/3}$ and
\begin{equation}
f_{eff}(\epsilon,\tilde{Z})=\sqrt{\frac{2\epsilon}{3}}\tilde{Z}^2-\frac{\tilde{Z}^3}{\sqrt{3}}\,.
\end{equation}
The height of the effective potential is $\ds
\frac{8}{27}\sqrt{\frac{2}{3}}\epsilon^{3/2}$ and the width is
$\sqrt{2\epsilon}$, see Fig.\ \ref{potential}.

As follows from the equations of motion, smallness of ${\epsilon}$
results in $|{\bm \nabla}Z| \sim p\epsilon \ll 1$. This allows one
to replace  $\left[1+({\bm \nabla} Z)^2\right]^{\frac{1}{2}}$ in
Eq.\ (\ref{eq1.1}) with $1 + \frac{1}{2}({\bm \nabla} Z)^2$.
Introducing dimensionless variables
\begin{eqnarray} \label{eq1.3}
&& x_0 =\left(\frac{2p\sqrt{\epsilon}}{3\sqrt{3}}\frac{\xi B_c^2}
{\eta a^2}\right)\tau,\quad
(x_1,x_2)=\left(\sqrt{2\epsilon/3}\;p\right)^{1/2}\frac{(x,y)}{a}
\nonumber \\ && v=V(x,y,Z)/\sigma p,\quad
u=\frac{3}{\sqrt{2\epsilon}}\left(Z/a-\tilde{Z}_{c}(1-\sqrt{2\epsilon/3})\right)
\nonumber \\
\end{eqnarray}
we obtain
\begin{eqnarray} \label{eq1.4}
S_{eff} & = &\frac{\sqrt{\epsilon}}{3\sqrt{6}\pi p}\frac{\eta
a^4}{\hbar}\oint\ud x_0\int\ud x_1\ud x_2 \Bigg[\frac{1}{2}({\bm
\nabla} u)^2
+u^2-\frac{u^3}{3} \nonumber \\
&+ & \frac{1}{2}\int_{\mathbf{R}}\ud
x_0'\frac{\left[u(x_0,x_1,x_2)-u(x_0',x_1,x_2)\right]^2}{(x_0-x_0')^2}\Bigg]
\end{eqnarray}
where $\nabla=(\partial_1,\partial_2)$.

\section{Instantons of the Dissipative 2+1 Model}\label{instanton}

Quantum depinning of the interface is given by the instanton
solution of the Euler-Lagrange equations of motion of the 2+1
field theory described by Eq.\ (\ref{eq1.4}):
\begin{equation}
\sum_{\mu = 0,1,2}\frac{\partial}{\partial
x^{\mu}}\left[\frac{\delta\mathcal{L}} {\delta\left({\partial
u}/{\partial x^{\mu}}\right)}\right]
-\frac{\partial\mathcal{L}}{\partial u} = 0\,.
\end{equation}
This gives
\begin{equation}
\label{eq1.7}
\nabla^2 u-2u+u^2-2\int_{\mathbf{R}}\ud x_0'\frac{u(x_0,x_1,x_2)-u(x_0',x_1,x_2)}
{(x_0-x_0')^2}=0
\end{equation}
with the boundary conditions
\begin{eqnarray}\label{eq1.8}
&&  u(-\Omega/2, x_1,x_2)=u(\Omega/2,x_1,x_2) \quad \forall
(x_1,x_2)\in \mathbf{R}^2 \nonumber \\
&& \max_{x_0\in[-\Omega/2,\Omega/2]} u(x_0,x_1,x_2)=u(0,x_1,x_2)
\quad\forall (x_1,x_2)\in \mathbf{R}^2 \nonumber \\
\end{eqnarray}
that must be periodic on imaginary time with the period
$\hbar/(k_BT)$. The corresponding period on $x_0$ is
\begin{equation}\label{Omega}
\Omega=\left(\frac{2p\sqrt{\epsilon}}{3\sqrt{3}}\frac{\xi
B_c^2}{\eta a^2}\right)\frac{\hbar}{k_{B}T}\,.
\end{equation}
This equation cannot be solved analytically, so we must proceed by
means of numerical methods.

\subsection{Zero temperature}
We apply the Fourier transform
\begin{equation}
\hat{u}(\vec{\omega})=\frac{1}{(2\pi)^{3/2}}\int_{\mathbf{R}^3}u(\vec{x})
e^{i\vec{\omega}\cdot\vec{x}}\ud^{3}x
\end{equation}
to equation \eqref{eq1.7} and get
\begin{equation}
\label{eq1.9}
\hat{u}(\vec{\omega})=\frac{(2\pi)^{-3/2}}{2+2\pi|\omega_0|+\omega_1^2+\omega_2^2}
\int_{\mathbf{R}^3}\ud^3\omega'\hat{u}
(\vec{\omega}-\vec{\omega}')\hat{u}(\vec{\omega}')
\end{equation}
which is still an integral equation for $\hat{u}(\vec{\omega})$.
The effective action \eqref{eq1.4} in terms of
$\hat{u}(\vec{\omega})$ becomes
\begin{eqnarray}\label{eq1.10}
&&S_{eff}\left[\hat{u}\right]=\frac{\sqrt{\epsilon}}{3\sqrt{6}\pi
p}\frac{\eta a^4}{\hbar}
\Bigg[\int_{\mathbf{R}^3}\ud^3\omega\;\hat{u}(\vec{\omega})\hat{u}(-\vec{\omega})\times
\nonumber \\
&& \Big(\frac{1}{2}(\omega_1^2+\omega_2^2)+1+\pi|\omega_0|\Big) - \nonumber \\
&&\frac{1}{3(2\pi)^{3/2}} \int_{\mathbf{R}^6}\ud^3
\omega\ud^3\omega'\hat{u}(\vec{\omega})\hat{u}(\vec{\omega}')\hat{u}
(-\vec{\omega}-\vec{\omega}')\Bigg].
\end{eqnarray}

We use the algorithm that is a field-theory extension of the
algorithm introduced in Refs. \onlinecite{Chang,Waxman} for the
problem of dissipative quantum tunneling of a particle. It
consists of the following steps:
\begin{enumerate}
 \item Start with an initial aproximation $\hat{u}_0(\vec{\omega})$.
Define the operator
 \begin{eqnarray}
   && \hat{O}:  \mathbf{R}\times\mathcal{L}^2(\mathbf{R}^3) \rightarrow
   \mathcal{L}^2(\mathbf{R}^3) \\
      && (\lambda,\hat{u}(\vec{\omega})) \mapsto \ds
      \frac{\lambda}{2+2\pi|\omega_0|+\omega_1^2+\omega_2^2}
\int_{\mathbf{R}^3}\ud^3\omega'\hat{u}(\vec{\omega}-\vec{\omega}')\hat{u}(\vec{\omega}')
\nonumber
 \end{eqnarray}
\item Let $\hat{u}_1(\vec{\omega})=\hat{O}(\lambda_0,\hat{u}_0(\vec{\omega}))$ for an initial
$\lambda_0\in\mathbf{R}$.
\item Calculate $\lambda_1=\lambda_0/\xi^{2}$ with $\xi=\frac{\hat{u}_1(\vec{\omega}=0)}
{\hat{u}_0(\vec{\omega}=0)}$.
\item Find $\hat{u}_2(\vec{\omega})=\hat{O}(\lambda_1,\hat{u}_1(\vec{\omega}))$.
\item Repeat steps $(2)-(4)$ until the successive difference satisfies a preset
convergence criterion.
\end{enumerate}
The output is the pair $(\lambda_n,\hat{u}_n(\vec{\omega}))$.
Finally, we apply a rescaling of $\hat{u}_n$ by a factor $\ds
(2\pi)^{3/2}\lambda_n$ to obtain the instanton solution. This
procedure leads to
\begin{equation}\label{WKB}
S_{eff} =\frac{\sqrt{\epsilon}}{3\sqrt{6}\pi p}\frac{\eta
a^4}{\hbar}I_0
\end{equation}
with numerical value of the integral $I_0 = 531\pm 19$. This
somewhat surprisingly large value of the integral has been
confirmed by our use of different computational grids.

\subsection{Non-zero temperature}
At $T \neq 0$ the period of the instanton solution is finite,
given by Eq.\ (\ref{Omega}). We look for a solution of the type
\begin{equation}
u(x_0,x_1,x_2)=\sum_{n\in\mathbf{Z}}
e^{i\omega_{0,n}x_0}u_n(x_1,x_2)
\end{equation}
with $\ds \omega_{0,n}={2\pi n}/{\Omega}$. Introducing into
\eqref{eq1.7} the above functional dependence and applying a 2D
Fourier transform we obtain
\begin{eqnarray}
\label{eq1.11}
&&\hat{u}_n(\vec{\omega})=\frac{1}{2+2\pi|\omega_{0,n}|+\vec{\omega}^2}
\times \nonumber \\
&&\left(\frac{1}{2\pi}\sum_{p\in\mathbf{Z}}\int_{\mathbf{R}^2}\ud^2\omega'\hat{u}_{n-p}
(\vec{\omega}-\vec{\omega}')\hat{u}_{p}(\vec{\omega}')\right)\,,
\end{eqnarray}
which is the integral equation for $\hat{u}_n$ with
$\vec{\omega}=(\omega_1,\omega_2)$. In terms of
$\big\{\hat{u}_{n}(\vec{\omega})\big\}_{n}$ the effective action
becomes
\begin{eqnarray}\label{eq1.12}
&&S_{eff}\left[\big\{\hat{u}\big\}_{n}\right]=\frac{\sqrt{\epsilon}}{3\sqrt{6}\pi
p}\frac{\eta a^4}{\hbar}\times \\
&&\Bigg[\sum_{n\in\mathbf{Z}}
\int_{\mathbf{R}^2}\ud^2\omega\hat{u}_{n}
(\vec{\omega})\hat{u}_{-n}(-\vec{\omega})
\left(\frac{\vec{\omega}^2}{2}+1+\pi|\omega_{0,n}|\right)-  \nonumber \\
&&\frac{1}{6\pi}\sum_{n,m\in\mathbf{Z}}\int_{\mathbf{R}^4}\ud^2
\omega\ud^2\omega'\hat{u}_{n}(\vec{\omega})\hat{u}_{m}(\vec{\omega}')
\hat{u}_{-n-m}(-\vec{\omega}-\vec{\omega}')\Bigg]\Omega \nonumber
\end{eqnarray}
The numerical algorithm is analogous to the one used in the $T=0$
case. It leads to
\begin{equation}
S_{eff} =\frac{\sqrt{\epsilon}}{3\sqrt{6}\pi p}\frac{\eta
a^4}{\hbar}I(T)
\end{equation}
The value of the integral depends on the value of $T$ in
comparison with the temperature, $T_c$, of the crossover from
quantum tunneling to thermal activation (see below). At $T \ll
T_c$ the numerical value of $I(T)$ is very close to $I_0$, while
at $T \gg T_c$ we recover the Boltzmann exponent, $S_{eff} =
V_{0}/(k_B T)$, with $V_{0}$ being the energy barrier for
depinning. Computation of $I(T)$ in the intermediate temperature
range requires very large computer time and will be reported
elsewhere. Nevertheless, as we shall see below, the crossover
temperature $T_c$ can be computed exactly.

\section{Crossover Temperature}\label{crossover}

The crossover temperature can be computed by means of theory of
phase transitions \cite{Tc}. Above $T_c$, the solution minimizing
the instantion action is a function
$u(x_0,x_1,x_2)=\bar{u}_{0}(x_1,x_2)$ that does not depend on
$x_0$. Just below $T_{c}$, the instanton solution can be split
into the sum of $\bar{u}_0$ and a term that depends $x_0$,
\begin{equation}
u(x_0,x_1,x_2)=\bar{u}_{0}(x_1,x_2)+u_{1}(x_1,x_2)\cos(2\pi/\Omega
x_0)\,.
\end{equation}
The instanton action is proportional to
\begin{eqnarray}\label{eq1.13}
\int_{\mathbb{R}^2} \ud x_1\ud x_2 \Phi(x_1,x_2; u,\nabla u)\,,
\end{eqnarray}
where  $\Phi(x_1,x_2; u,\nabla u)$ is the spatial action density.
Using the expansion of $u$ introduced in the previous section, we
obtain
\begin{eqnarray}
\label{eq1.14} &&\Phi(x_1,x_2;u_1,\nabla
u_1)=\Omega\left[\frac{1}{2}(\nabla
\bar{u}_0)^2+v(\bar{u}_0)\right]+ \nonumber \\
&& \frac{\Omega}{4}(\nabla u_1)^2 + \Lambda u_1^2+O(4)
\end{eqnarray}
with $\ds v(u)=u^2-{u^3}/{3}$ and
\begin{equation}\label{Lambda}
\Lambda = \frac{\Omega}{4}v''(\bar{u}_0)+\pi^2 \,.
\end{equation}

If $\Lambda > 0$, the only $(u_1,\nabla u_1)$ minimizing $\Phi$ is
$u_1\equiv0$, so we define the crossover temperature by the
equation
\begin{equation}\label{Lambda=0}
\min_{\vec{x}\in\mathbb{R}^2}\Lambda =
\min_{\vec{x}\in\mathbb{R}^2}\frac{\Omega_c}{4}v''[\bar{u}_0(x_1,x_2)]+\pi^2=0\,.
\end{equation}
Notice that this minimum corresponds to the minimum of
$v''[\bar{u}_0(x_1,x_2))]$. The equation of motion for $\bar{u}_0$
is
\begin{equation}
\label{eq1.15} \nabla^2 \bar{u}_0-2\bar{u}_0+\bar{u}_0^2=0\,.
\end{equation}
Solution corresponding to the minimum is spherically symmetric,
\begin{equation}
\bar{u}_0=\bar{u}_0\left(r=\sqrt{x_1^2+x_2^2}\right)\,,
\end{equation}
satisfying boundary conditions:  $\bar{u}_0\rightarrow0$ at
$r\rightarrow\infty$ and  $\bar{u}_0(0)=3$, which is the width of
the potential. Consequently,
\begin{eqnarray}
&&
\min_{\vec{x}\in\mathbb{R}^2}v''[\bar{u}_0(x_1,x_2)]=\min_{\bar{u}_0\in[0,3]}v''(\bar{u}_0)
\nonumber \\
&& =\min_{\bar{u}_0\in[0,3]}2(1-\bar{u}_0)=-4\,.
\end{eqnarray}
Then, according to equations (\ref{Lambda}) and (\ref{Lambda=0}),
the crossover temperature is determined by the equation $
\Omega(T_c) = \pi^2$, which gives
\begin{equation}\label{Tc-final}
T_c=\frac{2p\sqrt{\epsilon}}{3\sqrt{3}\pi^2}\frac{\hbar \xi
B_c^2}{k_{B}\eta a^2} =
\frac{4p\sqrt{\epsilon}}{3\pi^2\sqrt{3\kappa}} \frac{\hbar \rho_n
c^2}{k_B a^2} \,.
\end{equation}
\\

\section{Discussion}\label{discussion}

We are now in a position to discuss feasibility of the proposed
experiment on quantum depinning of the interface from a planar
defect in a type-I superconductor.  Two conditions must be
satisfied. Firstly the dimensionless effective action of Eq.\
(\ref{WKB}), which is the WKB exponent of the tunneling rate,
should not exceed $30-40$ in order for the tunneling to occur on a
reasonable time scale. Secondly, the crossover temperature
determined by Eq.\ (\ref{Tc-final}) better be not much less than
one kelvin. For a known superconductor, the two equations contain
three parameters: The parameter $p \leq 1$ describing the strength
of pinning, the parameter $a$ describing the width of the pinning
layer, and the parameter $\epsilon$ that controls how close the
transport current should be to the depinning current. We,
therefore, have to investigate how practical is the range of
values of these parameters that can provide conditions $S_{eff}
\sim 30$ and $T_c \sim 1$K.

Let us choose lead as an example. The values of $\lambda_L$ and
$\xi$ in lead are $37\,$nm and $38\,$nm, respectively, giving
$\kappa = \lambda_L/\xi = 0.45$. The critical field is $B_c
\approx 800\,$G. The elastic energy of the interface is $\sigma
\approx 0.4\,$erg/cm$^2$. The normal state resistivity in the
kelvin range is $5 \times 10^{-11}$ $\Omega\cdot$m = $5.6\times
10^{-21}$s., while the drag coefficient is $\eta \approx
0.35\,$erg$\cdot$s/cm$^4$. Then equations (\ref{WKB}) and
(\ref{Tc-final}) with conditions $S_{eff} \sim 30$ and $T_c \sim
1$K give $a/p^{1/3} \sim 3.7\,$nm and $\sqrt{\epsilon} a \sim
0.25\,$nm. If the pinning layer is incompatible with
superconductivity, then at $2a < \xi$ one should  expect $p \sim
2a/\xi$, giving $a \sim 1.65\,$nm and $\epsilon \sim 0.02$. This
means that observation of quantum escape of the interface from a
pinning layer of thickness $2a \sim 3.3\,$nm in a superconducting
Pb sample at $T \sim 1\,$K would require control of the transport
current within two percent of the critical depinning current. All
the above parameters are within experimental reach.

In conclusion, we have studied quantum escape from a planar
pinning defect of the interface separating superconducting and
normal regions in a type-I superconductor. This can correspond to
either quantum depinning of the interface from a grain boundary or
quantum depinning from an artificially prepared layer. The
computed tunneling rate, the required temperature and other
parameters all fall within realistic experimental range. We
encourage such experiment as it would present a rare opportunity
to study, in a controllable manner, dissipative quantum tunneling
of an extended object.

\section{Acknoweledgements}

The work at the University of Barcelona has been supported by the
Spanish Government project No. MAT2008-04535 and by Catalan ICREA
Academia. R.Z. acknowledges financial support from the Ministerio
de Ciencia e Innovaci\'{o}n de Espa\~na. The work of E.M.C. at
Lehman College has been supported by the U.S. Department of Energy
through grant No. DE-FG02-93ER45487.

\end{document}